\documentclass[sn-mathphys,Numbered]{sn-jnl}% American Physical Society (APS) Reference Style

\usepackage{graphicx}%
\usepackage{multirow}%
\usepackage{amsmath,amssymb,amsfonts}%
\usepackage{amsthm}%
\usepackage{mathrsfs}%
\usepackage[title]{appendix}%
\usepackage{xcolor}%
\usepackage{textcomp}%
\usepackage{manyfoot}%
\usepackage{booktabs}%
\usepackage{algorithm}%
\usepackage{algorithmicx}%
\usepackage{algpseudocode}%
\usepackage{listings}%
\usepackage{acronym}
\usepackage{enumitem}
\definecolor{chmagenta}{rgb}{0.54, 0.17, 0.88}

\acrodef{GW}{gravitational wave}
\acrodef{LIGO}{Laser Interferometer Gravitational-wave Observatory}
\acrodef{O1}{first observing run}
\acrodef{O2}{second observing run}
\acrodef{O3}{third observing run}
\acrodef{O4}{fourth observing run}
\acrodef{ML}{machine learning}
\acrodef{CNN}{convolutional neural network}
\acrodef{Conv}{convolutional}
\acrodef{FC}{fully-connected}

\begin{document}

\title[Gravity Spy: Lessons Learned]{Gravity Spy: Lessons Learned and a Path Forward}

%%% author list %%%
%%=============================================================%%
%% Prefix	-> \pfx{Dr}
%% GivenName	-> \fnm{Joergen W.}
%% Particle	-> \spfx{van der} -> surname prefix
%% FamilyName	-> \sur{Ploeg}
%% Suffix	-> \sfx{IV}
%% NatureName	-> \tanm{Poet Laureate} -> Title after name
%% Degrees	-> \dgr{MSc, PhD}
%% \author*[1,2]{\pfx{Dr} \fnm{Joergen W.} \spfx{van der} \sur{Ploeg} \sfx{IV} \tanm{Poet Laureate} 
%%                 \dgr{MSc, PhD}}\email{iauthor@gmail.com}
%%=============================================================%%

\author*[1,2,3]{\fnm{Michael} \sur{Zevin}}\email{mzevin@adlerplanetarium.org}
\equalcont{NASA Hubble Fellow}
%\equalcont{These authors contributed equally to this work.}

\author[4]{\fnm{Corey B.} \sur{Jackson}}
%\equalcont{These authors contributed equally to this work.}

\author[5]{\fnm{Zoheyr} \sur{Doctor}}
\author[6]{\fnm{Yunan} \sur{Wu}}
\author[7]{\fnm{Carsten} \sur{{\O}sterlund}}
\author[1,5]{\fnm{L. Clifton} \sur{Johnson}}
\author[8]{\fnm{Christopher P. L.} \sur{Berry}}
\author[7]{\fnm{Kevin} \sur{Crowston}}
\author[5]{\fnm{Scott B.} \sur{Coughlin}}
\author[5,9]{\fnm{Vicky} \sur{Kalogera}}  % needs dept of astronomy as well, 2145 sheridan road

\author[5]{\fnm{Sharan} \sur{Banagiri}}
\author[10]{\fnm{Derek} \sur{Davis}}
\author[11]{\fnm{Jane} \sur{Glanzer}}
\author[6]{\fnm{Renzhi} \sur{Hao}}
\author[6,5]{\fnm{Aggelos K.} \sur{Katsaggelos}}
\author[12]{\fnm{Oli} \sur{Patane}}
\author[5]{\fnm{Jennifer} \sur{Sanchez}}
\author[13]{\fnm{Joshua} \sur{Smith}}
\author[14]{\fnm{Siddharth} \sur{Soni}}
\author[1]{\fnm{Laura} \sur{Trouille}}
\author[15]{\fnm{Marissa} \sur{Walker}}

\author[16,21]{\fnm{Irina} \sur{Aerith}}
\author[17,21]{\fnm{Wilfried} \sur{Domainko}}
\author[18,21]{\fnm{Victor-Georges} \sur{Baranowski}}
\author[19,21]{\fnm{Gerhard} \sur{Niklasch}}
\author[20,21]{\fnm{Barbara} \sur{T\'egl\'as}}

\affil[1]{\orgdiv{Zooniverse}, \orgname{The Adler Planetarium}, \orgaddress{\street{1300 South DuSable Lake Shore Drive}, \city{Chicago}, \postcode{60605}, \state{IL}, \country{USA}}}

\affil[2]{\orgdiv{Kavli Institute for Cosmological Physics}, \orgname{The University of Chicago}, \orgaddress{\street{5640 South Ellis Avenue}, \city{Chicago}, \postcode{60637}, \state{IL}, \country{USA}}}

\affil[3]{\orgdiv{Enrico Fermi Institute}, \orgname{The University of Chicago}, \orgaddress{\street{933 East 56th Street}, \city{Chicago}, \postcode{60637}, \state{IL}, \country{USA}}}

\affil[4]{\orgdiv{Information School}, \orgname{University of Wisconsin–Madison}, \orgaddress{\street{600 N
Park Street}, \city{Madison}, \postcode{53706}, \state{WI}, \country{USA}}}

\affil[5]{\orgdiv{Center for Interdisciplinary Exploration and Research in Astrophysics (CIERA)}, \orgname{Northwestern University}, \orgaddress{\street{1800 Sherman Ave}, \city{Evanston}, \postcode{60201}, \state{IL}, \country{USA}}}

\affil[6]{\orgdiv{Department of Electrical and Computer Engineering}, \orgname{Northwestern University}, \orgaddress{\street{2145 Sheridan
Road}, \city{Evanston}, \postcode{60208}, \state{IL}, \country{USA}}}

\affil[7]{\orgdiv{School of Information Studies}, \orgname{Syracuse University}, \orgaddress{\street{343 Hinds Hall}, \city{Syracuse}, \postcode{13210}, \state{NY}, \country{USA}}}

\affil[8]{\orgdiv{SUPA, School of Physics and Astronomy}, \orgname{University of Glasgow}, \orgaddress{\street{Kelvin Building,
University Ave}, \city{Glasgow}, \postcode{8QQ}, \state{G12}, \country{UK}}}

\affil[9]{\orgdiv{Department of Physics and Astronomy}, \orgname{Northwestern University}, \orgaddress{\street{2145 Sheridan Road}, \city{Evanston}, \postcode{60208}, \state{IL}, \country{USA}}}

\affil[10]{\orgdiv{LIGO Laboratory}, \orgname{California Institute of Technology}, \orgaddress{\street{1200 East California Boulavard}, \city{Pasadena}, \postcode{91125}, \state{CA}, \country{USA}}}

\affil[11]{\orgdiv{Department of Physics and Astronomy}, \orgname{Louisiana State
University}, \orgaddress{\street{202 Nicholson Hall}, \city{Baton Rouge}, \postcode{70803}, \state{LA}, \country{USA}}}

\affil[12]{\orgname{LIGO Hanford Observatory}, \orgaddress{\street{127124 N Route 10}, \city{Hanford}, \postcode{99354}, \state{WA}, \country{USA}}}

\affil[13]{\orgdiv{The Nicholas and Lee Begovich Center for Gravitational-Wave Physics and Astronomy}, \orgname{California State University Fullerton}, \orgaddress{\street{800 N. State College Blvd}, \city{Fullerton}, \postcode{92831}, \state{CA}, \country{USA}}}

\affil[14]{\orgdiv{LIGO Laboratory}, \orgname{Massachusetts Institute of Technology}, \orgaddress{\street{185 Albany Street}, \city{Cambridge}, \postcode{02139}, \state{MA}, \country{USA}}}

\affil[15]{\orgdiv{Department of Physics, Computer Science and Engineering}, \orgname{Christopher Newport University}, \orgaddress{\street{One Avenue of the Arts}, \city{Newport News}, \postcode{23606}, \state{VA}, \country{USA}}}

\affil[16]{\city{Cedar Falls}, \state{IA}, \country{USA}}

\affil[17]{\city{Hermagor}, \country{Austria}}

\affil[18]{\orgname{Sorbonne Paris Nord University}, \orgaddress{\street{99 Avenue Jean Baptiste Cl\'ement}, \city{Villetaneuse}, \postcode{93430}, \country{France}}}

\affil[19]{\orgname{ConSol Software GmbH}, \orgaddress{\street{St.-Cajetan-Strasse 43}, \city{Munich}, \postcode{81669}, \country{Germany}}}

\affil[20]{\city{Budapest}, \country{Hungary}}

\affil[21]{\orgname{Gravity Spy Moderator}}

%%% abstract %%%
\abstract{
The Gravity Spy project aims to uncover the origins of glitches, transient bursts of noise that hamper analysis of gravitational-wave data. 
By using both the work of citizen-science volunteers and machine-learning algorithms, the Gravity Spy project enables reliable classification of glitches. 
Citizen science and machine learning are intrinsically coupled within the Gravity Spy framework, with machine-learning classifications providing a rapid first-pass classification of the dataset and enabling tiered volunteer training, and volunteer-based classifications verifying the machine classifications, bolstering the machine-learning training set and identifying new morphological classes of glitches. 
These classifications are now routinely used in studies characterizing the performance of the LIGO gravitational-wave detectors. 
Providing the volunteers with a training framework that teaches them to classify a wide range of glitches, as well as additional tools to aid their investigations of interesting glitches, empowers them to make discoveries of new classes of glitches. 
This demonstrates that, when giving suitable support, volunteers can go beyond simple classification tasks to identify new features in data at a level comparable to domain experts. 
The Gravity Spy project is now providing volunteers with more complicated data that includes auxiliary monitors of the detector to identify the root cause of glitches.}

%%% keywords %%%
\keywords{citizen science, LIGO, detector characterization, machine learning}

\maketitle

%%% main text %%%
\section{Introduction}\label{sec1}
\Acp{GW}, ripples in the fabric of space and time that are a key prediction of Albert Einstein's century-old theory of general relativity~\cite{Einstein:1916cc,Einstein:1918btx}, were directly observed for the first time by the Advanced \ac{LIGO} instruments in September 2015~\citep{Abbott:2016blz}. 
Since this groundbreaking discovery, the two \ac{LIGO} detectors~\cite{TheLIGOScientific:2014jea} and their partner, the Virgo detector~\cite{TheVirgo:2014hva} have measured nearly $100$ \ac{GW} candidates from coalescing black holes and neutron stars during their first three observing runs~\cite{LIGOScientific:2021djp}. 
The \ac{O4} of the \ac{LIGO}, Virgo, and now KAGRA~\cite{KAGRA:2018plz} \ac{GW} detector network is under way and expected to result in hundreds more \ac{GW} detections over the next two years~\cite{Abbott:2020qfu}. 

\begin{figure}
\centering
\includegraphics[width=\linewidth]{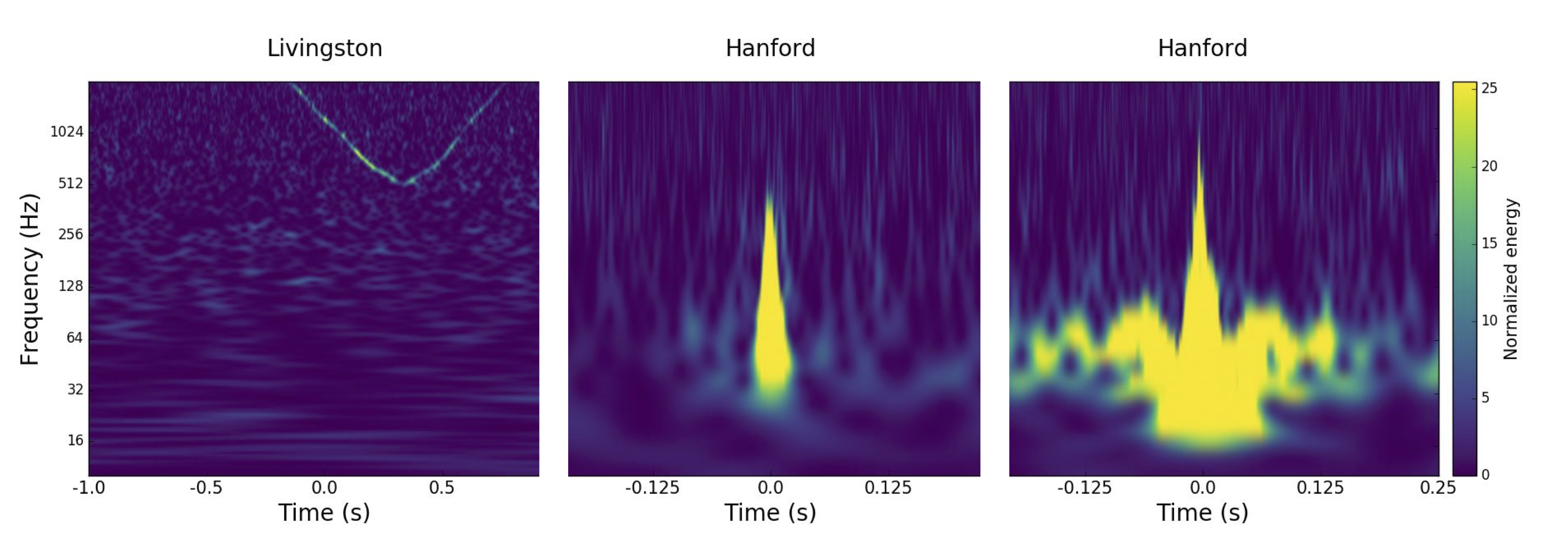}
\caption{Spectrograms of example glitches seen in \ac{LIGO} detectors. 
Time is along the horizontal axis, and frequency along the vertical axis. 
The color denotes the energy in each time-frequency bin. 
These glitches exemplify a subset of glitch classes. 
The left panel shows a Whistle glitch, middle panel a Blip glitch, and right panel a Koi Fish glitch.}
\label{fig:egGlitches}
\end{figure}

To observe \acp{GW} from ground-based detectors such as \ac{LIGO} and Virgo, the instruments need to be sensitive to changes in the length of the detector arms on the order of $10^{-18}~\mathrm{m}$~\cite{LIGOScientific:2019hgc}. 
State-of-the-art instrumentation and data analysis techniques have been developed to achieve these sensitivity requirements. 
However, non-astrophysical noise sources, originating from both instrumental and environmental factors, affect the detectors and lessen their sensitivity to the \ac{GW} universe. 
Of particular concern are transient, non-Gaussian noise sources known as \emph{glitches} that appear in the detectors at a high rate \cite{Blackburn:2008ah,TheLIGOScientific:2016zmo,Davis:2021ecd}. 
Though some glitches have known causes, others show no obvious correlation with instrumental and environmental noise monitors, making their root causes difficult to diagnose~\cite{Nuttall:2018xhi,Davis:2022dnd}. 
The classification and characterization of glitches are paramount to minimizing their effect on \ac{GW} measurements. Classification is the crucial first step because many glitches are morphologically similar, as can be identified through their spectrograms. 
Figure~\ref{fig:egGlitches} shows spectrograms of three glitches that exemplify common glitch classes seen in the \ac{LIGO} detectors: Whistles, Blips and Koi Fish~\cite{Zevin:2016qwy,Glanzer:2022avx}.   
Although these glitches can be differentiated by eye in their spectrograms (and by their sound, if played through a speaker), classification is a daunting task because of the sheer volume of glitch data that is accumulated throughout an observing run. 

Such large-scale data analysis challenges are not unique to \ac{GW} astronomy. Fundamental challenges and opportunities for 21st-century science lie in developing machinery and techniques to characterize and use expansive data. 
The paradigm in which individual researchers or even teams of researchers can analyze data has shifted to relying on novel methods for analysis of large-scale scientific datasets. 

\Ac{ML} techniques are heavily embedded in scientific data analysis, becoming a standard approach for analyzing large-scale data \cite[e.g.,][]{Schmidt2019npjCM...5...83S, Ball:2009wd}; however, there are many instances where analysis by humans is still critical. 
One example is identifying new data classes not initially accounted for in training sets for \ac{ML} algorithms. 
This novelty is indeed the case for \ac{GW} detector characterization, as new glitches regularly appear due to environmental and instrumental changes as the detectors evolve~\cite{Davis:2022dnd,Soni:2021cjy}. 
To this end, the Gravity Spy project was created to characterize glitches in \ac{GW} detectors by combining computer- and human-based classification schemes~\cite{Zevin:2016qwy}. 

At its core, Gravity Spy is a citizen-science project.%
\footnote{Gravity Spy Project \href{https://gravityspy.org}{www.gravityspy.org}.}
As a citizen-science project, it involves members of the general public in the scientific process, including formulating research questions, collecting or analyzing data, observing and recording natural phenomena, and disseminating results~\cite{Bonney:2009et}. 
As Internet-enabled devices have become increasingly ubiquitous, citizen-science projects have become a feasible approach to providing access to human insights at a large scale. 
For example, Galaxy Zoo~\cite{Lintott:2008ne} invites volunteers to engage in the morphological classification of images of galaxies produced by the Sloan Digital Sky Survey: the project has successfully increased engagement with the public and led to novel discoveries in astronomy.%
\footnote{Galaxy Zoo Results \href{https://www.zooniverse.org/projects/zookeeper/galaxy-zoo/about/results}{www.zooniverse.org/projects/zookeeper/galaxy-zoo/about/results}.}

Gravity Spy has a web-based interface through which anyone can provide analysis of \ac{LIGO} glitches and interact with \ac{LIGO} scientists about the state of the detectors.
It is hosted on the Zooniverse citizen-science platform.%
\footnote{Zooniverse \href{https://www.zooniverse.org/}{www.zooniverse.org}.}
The Zooniverse platform has fielded a workable crowd-sourcing model (involving over $2.5$ million people in more than four hundred projects since its inception) through which volunteers provide large-scale scientific data analysis. 
Zooniverse provides tools for systematically performing data-analysis tasks on data collections, making it an ideal platform for Gravity Spy.

Building a citizen-science project using Zooniverse would not solve the specific challenges \ac{GW} detectors face alone. 
The glitches in the detectors can change over the course of a run as \ac{GW} scientists adjust the detectors or from other transient terrestrial events. 
Understanding and mitigating glitches requires innovative solutions, testing, and meticulous analysis of the data that only an interdisciplinary team of citizen-science volunteers, \ac{LIGO} detector-characterization experts, computer scientists trained in machine learning, and social scientists could provide. 

Since its inception in October 2016, Gravity Spy has accumulated over 7 million classifications of glitches by tens of thousands of Zooniverse users, significantly contributing to the characterization and identification of glitches in \ac{LIGO} data. 
Gravity Spy also provides a testbed for fostering a symbiotic relationship between human- and machine-classification approaches. 
This connection goes far deeper than the classification task itself; the interplay between the human and machine components of the project allows for a customized training regimen for project volunteers, a more efficient and individually-tailored classification task, and the ability of volunteers to actively improve the underlying \ac{ML} algorithms. 

This article details activities undertaken as part of the three-year multi-institution National Science Foundation grant (INSPIRE 1547880) to develop Gravity Spy as a prototype for the next generation of citizen-science projects. 
We describe the Gravity Spy project, its novel approaches to citizen science, its impact on \ac{GW} detector characterization over the past half-decade, and the citizen-science techniques exclusively developed for this project. 
We will argue why such techniques are necessary for citizen science to remain highly relevant even as datasets grow exponentially and machine-based classification algorithms advance. 
Following an overview of the project, results, and lessons learned, we will look ahead to future advancements and adaptations to the Gravity Spy project, particularly focusing on the latest extension known as Gravity Spy 2.0. 

\begin{figure}
\centering
\includegraphics[width=\linewidth]{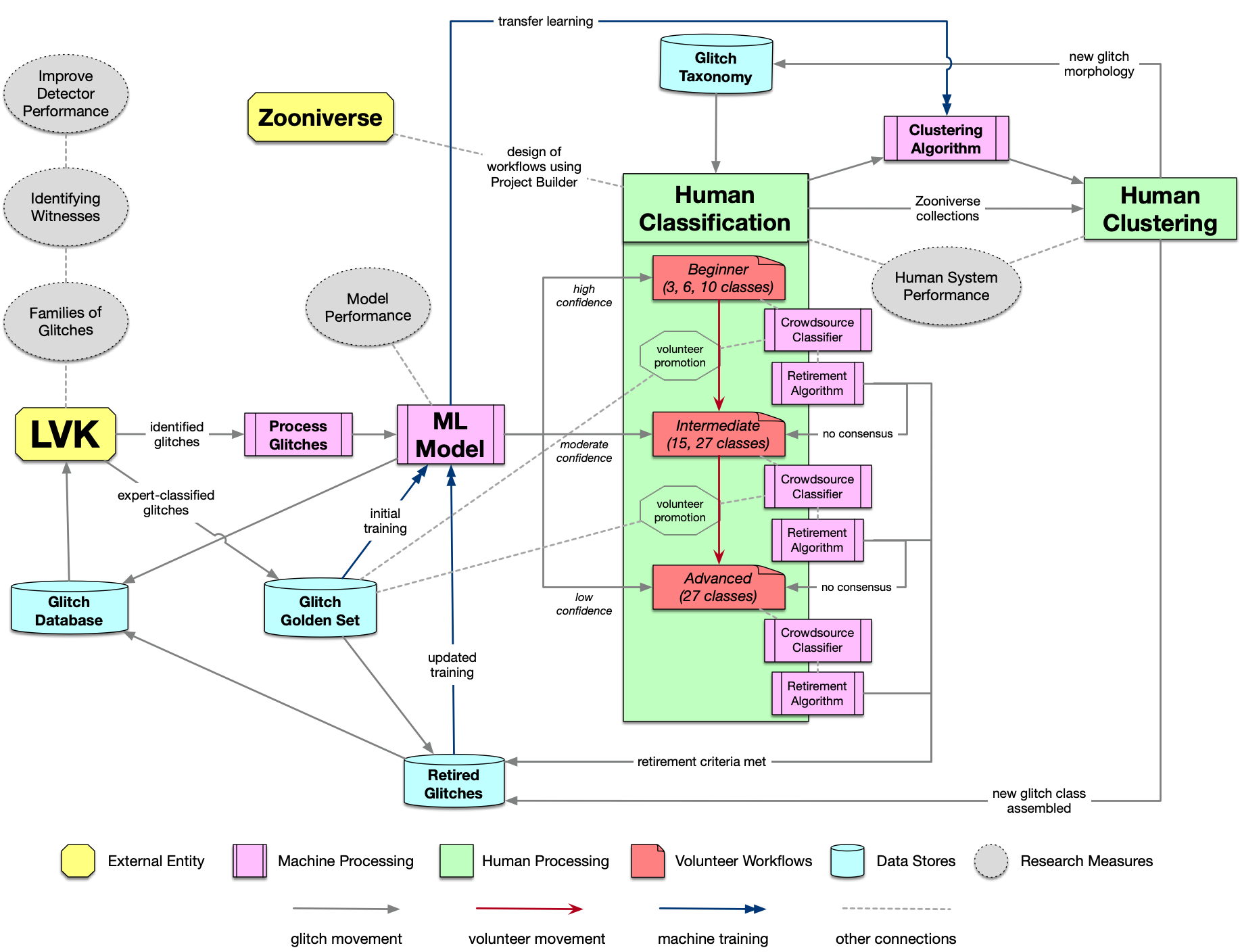}
\caption{Components of the interconnected Gravity Spy system. 
Gray arrows show the movement of glitches throughout the project, red arrows the progression of volunteers, and blue arrows the training of \ac{ML} models. 
We note that there are multiple levels that volunteers progress through in the Beginner and Intermediate workflows with differing number of glitch classes, though the \ac{ML} confidence of glitches are consistent within all Beginner workflows and within all Intermediate workflows. 
}
\label{fig:scheme}
\end{figure}

\section{Building Gravity Spy}\label{sec2}

The Gravity Spy project relies on an intricate interconnection between \ac{GW} data analysis, citizen science, and machine learning. 
An overview of the main Gravity Spy system is shown in Figure~\ref{fig:scheme} and summarized below: 
\begin{enumerate}
    \item Glitches are identified using the \texttt{Omicron} pipeline~\cite{Robinet:2020lbf}, which identifies moments with excess power in the data stream of each \ac{LIGO} detector individually, referred to as \texttt{Omicron} \textit{triggers}. 
    \texttt{Omicron} triggers that exceed a signal-to-noise threshold and are from times of a suitable detector state are selected for study. 
    \item Spectrograms like those in Figure~\ref{fig:egGlitches} are created for four different time durations around each trigger. 
    Each particular trigger's collection of spectrograms in the project is referred to as a \emph{subject}. 
    \item Each new subject is sent through a trained \ac{ML} algorithm and assigned a probability of being an instance of one of the pre-determined morphological classes in the Gravity Spy project. 
    \item New subjects are distributed to the Gravity Spy volunteer workflows based on the confidence score from the initial \ac{ML} classification, with glitches likely to be of a known glitch class being provided to new volunteers and less certain glitches to more advanced volunteers. 
    \item Volunteers morphologically classify subjects until the combined machine- and human-classification score reaches a predetermined threshold, at which point that particular subject is retired from the project (i.e., removed from the classification workflows). 
    If no consensus is reached after a set number of classifications, the glitch is migrated to a more advanced workflow. 
    \item Volunteers advance through the project and gain the ability to access workflows with additional classes of glitches as they correctly classify \emph{golden} subjects (i.e., subjects that experts have already classified).%
    \footnote{The original plan for Gravity Spy to use the unclassified subjects with high-confidence \ac{ML} scores in advancing volunteers through the project as to not waste volunteer classifications on glitches with already-known labels; however, this proved to be suboptimal due to the \ac{ML} algorithm spuriously identifying certain glitches incorrectly with high confidence.}  
    \item Volunteers in the most advanced workflows examine glitches that neither human nor machine classification have been able to confidently identify, in order to look for possible new classes of glitches that can be proposed for follow-up by \ac{LIGO} scientists. 
    \item All machine-learning, volunteer, and combined classification results for active and retired subjects are provided to LIGO--Virgo--KAGRA (LVK) scientists to aid glitch studies. 
    Retired images are actively added to the \ac{ML} training set, which can be retrained at a predetermined latency to improve its initial classification ability.%
    \footnote{The active addition of retired subjects to the \ac{ML} training set and model retraining has yet to be fully automated.} 
\end{enumerate}
The following subsections provide more details about the building and performance of each component of the Gravity Spy system. 

\subsection{The Dataset: Transient Noise in \ac{GW} Detectors}\label{subsec:2.1}

As summarized above, Gravity Spy classifications operate on spectrograms of \ac{GW} data around the times of \texttt{Omicron} triggers \cite{Robinet:2020lbf}. 
The \texttt{Omicron} software relies on the $Q$-transform, an analog of the short Fourier transform~\cite{Chatterji:2004qg}. 
Times of excess power are identified through the $Q$-transform, and the $Q$-transformed data are represented as spectrogram images (also known as $Q$-scans, see Figure~\ref{fig:egGlitches}). 

A glitch in one of the \ac{LIGO} detectors will trigger \texttt{Omicron} to produce spectrograms for four different time windows: $0.5~\mathrm{s}$, $1.0~\mathrm{s}$, $2.0~\mathrm{s}$, and $4.0~\mathrm{s}$~\cite{Zevin:2016qwy}.  
The reason multiple time windows are used is so that both humans and \ac{ML} models can examine glitch morphologies that occur on different characteristic timescales. 
These images are then exported to an \ac{ML} model and for human vetting, as described in later sections of this article. 
The classifications from the \ac{ML} and humans are then added to the Gravity Spy dataset for use by analysts. 

Since the start of Advanced \ac{LIGO} observations in the fall of 2015 through the end of the \ac{O3} in spring 2020, the Gravity Spy dataset has accumulated $\gtrsim 1.4 \times 10^6$ glitch triggers from the \texttt{Omicron} pipeline with a signal-to-noise ratio $>7.5$, corresponding on average to a new glitch being added to the project per every minute of observing time. 
Compared to the $\sim 10^2$ astrophysical events observed over this timespan~\cite{LIGOScientific:2021djp}, the number of glitches that have been uploaded to Gravity Spy is $>10^4$ times larger. 
As of the start of the fourth observing run of the Advanced \ac{LIGO}, Advanced Virgo and KAGRA detectors, $27$ morphological classes~\cite{Glanzer:2022avx}, including the catch-all None-of-the-Above (NOTA) and No Glitch classes are available for volunteer classification of \ac{LIGO} glitches, with $23$ incorporated into the \ac{ML} classifier~\cite{Soni:2021cjy}.

\subsection{The Gravity Spy Classification Task}\label{subsec:2.2}

The Gravity Spy classification task on the Zooniverse user interface involves citizen-science volunteers reviewing images of glitches and classifying them into morphological categories. 
The different levels at which volunteers can classify glitches are called \emph{workflows}.  
As volunteers learn the classification task and classify glitches in accordance with glitches pre-classified by experts, they unlock and advance through workflows that introduce more glitch classes, more options in the classification interface, and glitches that are classified with less confidence by the \ac{ML} algorithm. 

\begin{figure}
\centering
\includegraphics[width=\linewidth]{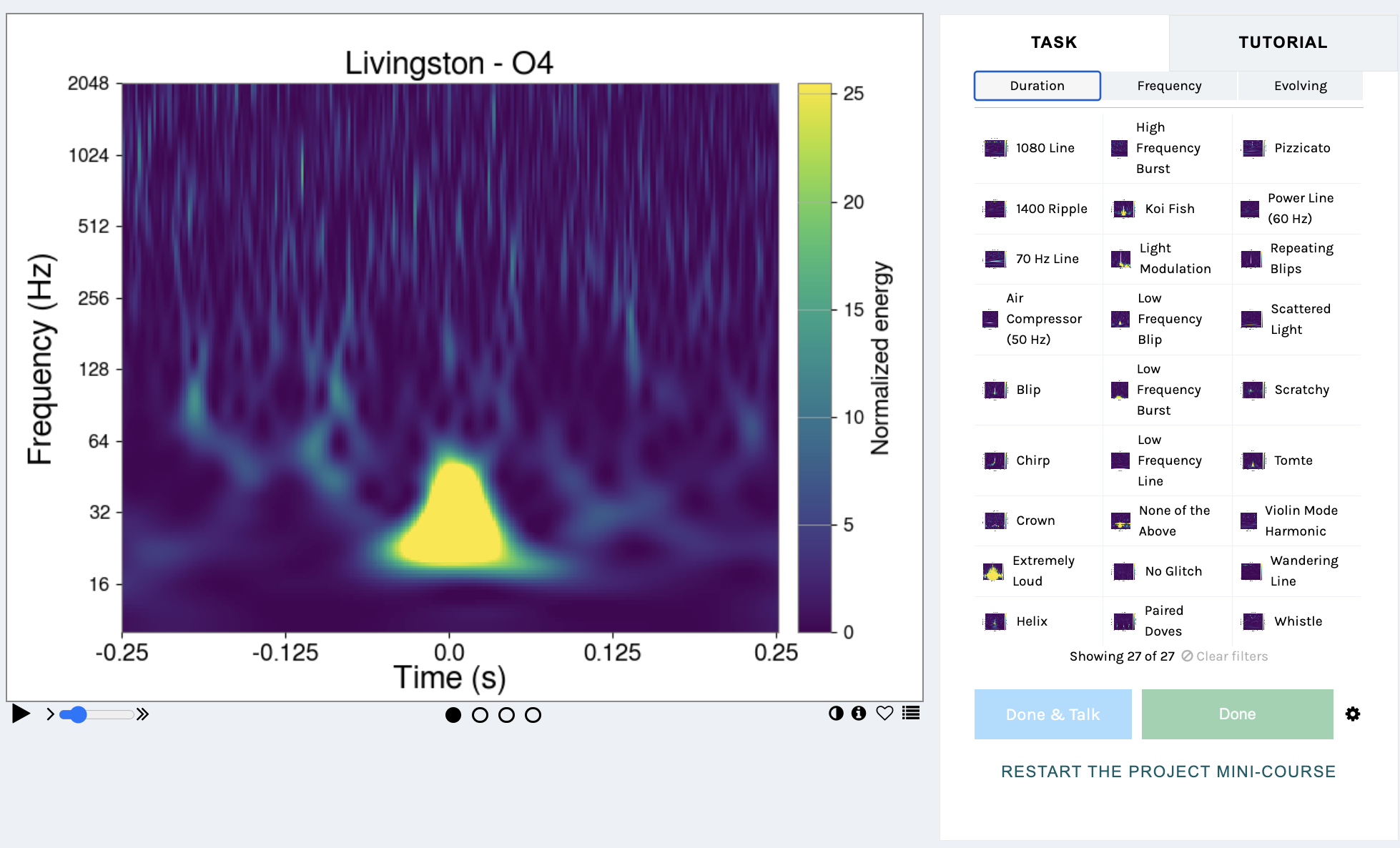}
\caption{The Gravity Spy classification interface for advanced workflows, named (\texttt{Binary Black Hole Merger (Level 5)} and \texttt{Inflationary Gravitational Waves (Level 6)}). 
These workflows allow for volunteers to pick from all of the $24$ morphological classes that are currently accounted for in the Gravity Spy project. }
\label{fig:interface}
\end{figure}

The classification interface for the most advanced workflow is shown in Figure~\ref{fig:interface}. 
The left side of the interface presents the glitch to be classified, and the right side, a list of possible classes, along with NOTA for glitches that do not match any listed classes. 
To help volunteers pick a class, a Field Guide with exemplary images and text descriptions is available. 
Metadata (e.g., the date on which the glitch occurred) and image filtering options are present in the interface. 
Volunteers can also save favorite subjects or create collections of subjects that they can view after the glitch has been classified. 

Upon selecting a class, volunteers are presented with two options: Done or Done \& Talk. 
Both options submit the response to the response database. 
If the subject was used to evaluate promotion (gold-standard data), feedback is provided that the volunteer agreed or disagreed with the expert evaluation of the glitch. 
Selecting Done \& Talk directs the volunteer to a thread on the Gravity Spy Talk discussion board where they can review comments posted about the image by other volunteers or post a new comment.%
\footnote{Gravity Spy Talk \href{https://www.zooniverse.org/projects/zooniverse/gravity-spy/talk}{www.zooniverse.org/projects/zooniverse/gravity-spy/talk}.}
The Talk feature is extensively used to comment on NOTA glitches that might represent a new glitch class.

\subsection{Machine Learning for Gravity Spy: Classification and Volunteer Training}\label{subsec:2.3}

\ac{ML} processing is an integral part of the Gravity Spy architecture and is used for classification and volunteer training. 
The \ac{ML} model used for initial classification is a \ac{CNN}, a class of deep learning algorithms that shows exceptional performance in image recognition endeavors~\cite{krizhevsky2012imagenet}. 
The detailed architecture of the \ac{CNN} used in the Gravity Spy classification task and how glitch spectrograms with differing temporal durations are combined can be found in \cite{Zevin:2016qwy,Bahaadini:2017dqg}. 

A training set of $\simeq 7700$ labeled glitches across the $19$ initial morphological classes was synthesized for the initial Gravity Spy launch, and this training set has been supplemented over time to include nearly $10^4$ labeled glitches over $23$ classes after \ac{O3}~\cite{Glanzer:2022avx}. 
This dataset continues to be expanded as new data are being taken. 
The initial training set was constructed by experts: first, $\sim 1000$ labeled glitches were used to train a preliminary \ac{ML} model (with relatively poor accuracy), and then these \ac{ML} labels were vetted by experts to expand the initial training set. 

Gravity Spy uses a novel approach for volunteer training and advancement that relies on \ac{ML} to provide task scaffolding. 
The design is informed by learning theories, namely the zone of proximal development~\cite{vygotsky1978MindInSociety,engestrom2014learning, jackson2020teaching}. 
Specifically, new volunteers to the project progress through a series of increasingly challenging and complex workflows to develop their knowledge of the classification task and the zoo of glitch morphologies, with more advanced workflows and more difficult-to-classify glitches opening to volunteers as they proceed through the project.
The \ac{ML} classifications and confidence scores determine the workflow to which a particular subject is assigned. 
Subjects that the \ac{ML} confidently classified are assigned to beginner workflows, while subjects that the \ac{ML} is uncertain about, which may thus represent new categories of glitches, are assigned to the more advanced workflows. 
There are multiple beginner workflows containing an increasing number of glitch classes. 
Glitches are assigned to the appropriate workflows so that new volunteers see glitches that are likely (but not certain) to be examples of one of the included classes. 
In the initial level, volunteers begin with only two options and, upon meeting performance thresholds as assessed by classification of gold-standard data, are promoted to subsequent levels in which additional glitch classes are presented. 
Also, instances of the glitch classes introduced to the volunteer in the previous level may be more challenging to classify, as indicated by lower \ac{ML} confidence scores. 
At each level, volunteers have the option of NOTA to handle the case where the \ac{ML} has confidently misclassified a glitch. 
In the most advanced workflow, most glitches have low \ac{ML} confidence and may be representative of new morphological classes. 
The task of the volunteers thus shifts from classifying glitches into known classes to searching for similarities in the images that indicate a possible novel class of glitch. 

As volunteers classify images, their classifications (which are coupled to a unique \emph{confusion matrix} for each user that determines which classes a particular user commonly confuses with another) are combined with the initial \ac{ML} confidence score to determine whether a particular subject has been classified with a high enough pre-set accuracy to be retired from the project and added to the \ac{ML} training set; these glitches that are added to the \ac{ML} training set improve its morphological coverage. 
Theoretically, this iterative retraining of the \ac{ML} model could be automated. 
However, several variables govern this retirement procedure, and investigation into the optimal settings for these retirement criteria is ongoing.

\subsection{Volunteers of Gravity Spy: Supporting the Machine and Making Discoveries}

Hand-labeling glitches with the predefined glitch classes represents the lion’s share of work in Gravity Spy, without which the \ac{ML} could not be trained nor benchmarked. 
Gravity Spy volunteers are essential in providing these by-eye classifications. 
As discussed earlier, the input provided by volunteers goes beyond the initial classification task; volunteers help identify and characterize glitches that do not fit into previously known glitch classes. 
When glitches do not fit one of the known glitch classes in the primary labeling, volunteers can begin by labeling them as NOTA, and proceed to create collections of such glitches that exhibit similar morphologies. 

\begin{figure}
\centering
\includegraphics[width=0.5\linewidth]{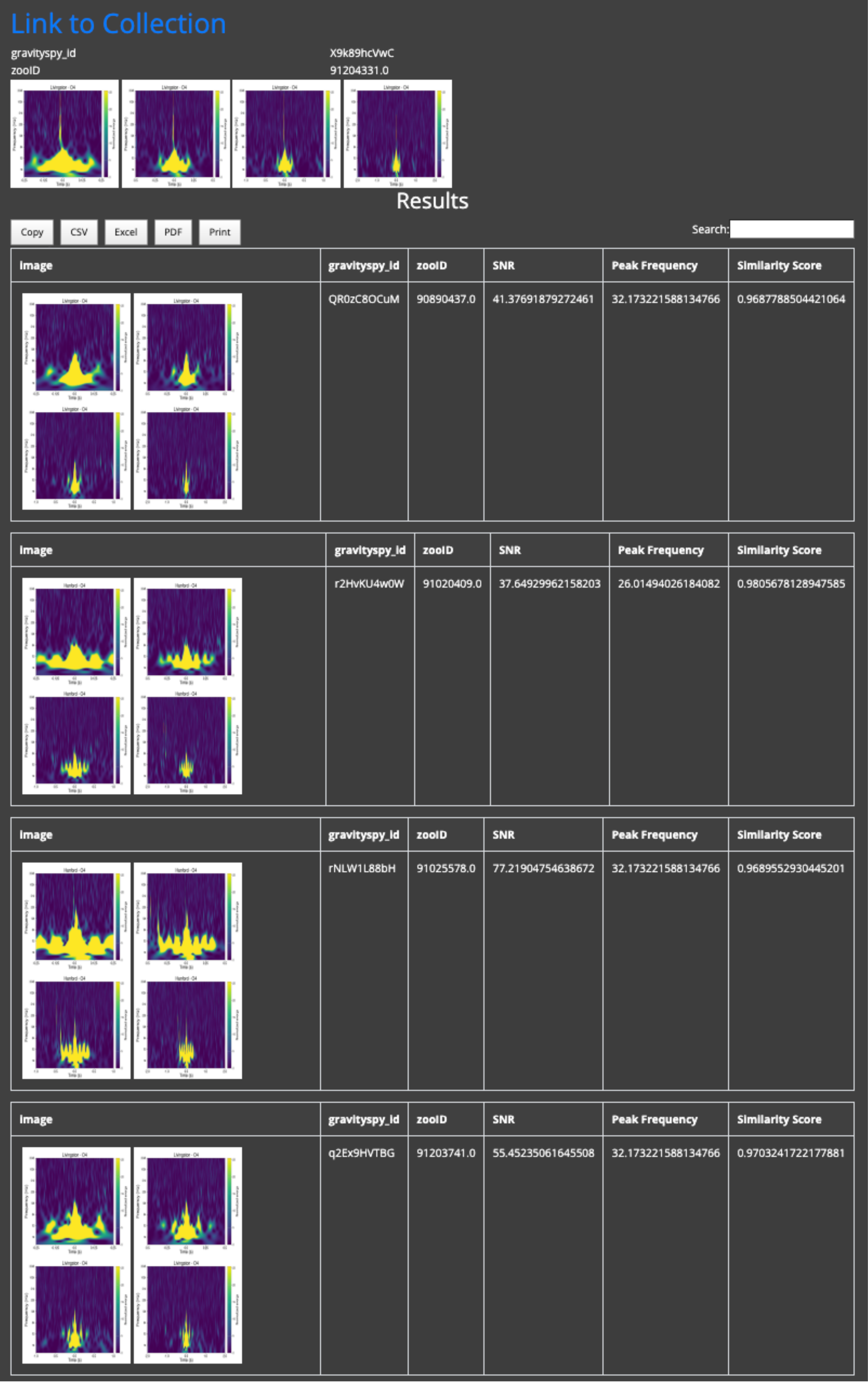} 
\caption{Results from a query of the Similarity Search tool. 
The glitch at the top of the image is was glitch that was queried on, and the four glitches below are some of the most similar glitches in the database. 
Similarity scores are in the rightmost column of the table. }
\label{fig:simp}
\end{figure}

Large numbers in a new glitch class indicate that this morphology may be particularly detrimental to \ac{GW} detector sensitivity. 
Collecting large samples of such glitches can allow for \ac{LIGO} scientists to identify trends (e.g., in the times that they occur or in the auxiliary sensors that are triggered at the time of the glitch)~\cite{Cabero:2019orq,Davis:2022dnd}. 
In addition to supplementing glitch collections by continuing through the classification task and identifying similar images, which can be cumbersome and time consuming, we enable volunteers to perform additional data analysis tasks using additional tools side by side with Zooniverse, collectively known as \textit{Gravity Spy Tools}.\footnote{Gravity Spy Tools \href{https://www.gravityspytools.ciera.northwestern.edu}{gravityspytools.ciera.northwestern.edu}.} 
One such tool that has proved to be highly useful is the \textit{Similarity Search}, in which volunteers (and project scientists) are able to query similar-looking glitches within the full dataset.\footnote{Gravity Spy Similarity Search \href{https://gravityspytools.ciera.northwestern.edu/search/}{gravityspytools.ciera.northwestern.edu/search/}.}
The search utilizes transfer learning, first modeling the properties of existing glitch classes in a high-dimensional feature space and then relying on a clustering algorithm to identify images that are morphologically similar in the database~\cite{Bahaadini:2018nkh,Coughlin:2019ref}. 
To use the tool, project scientists or volunteers input a particular glitch subject (each glitch has a unique subject label), and query other glitches in the dataset that most closely resemble that subject morphologically. 
After running the Similarity Search (a screenshot of the output is shown in Figure~\ref{fig:simp}), users can evaluate the metadata of resulting glitches, decide which images to include or exclude, and export the results of the search to a new collection. 
The Similarity Search plays a crucial role by effectively filtering out the majority of the non-matching glitches, significantly enhancing the purity of the set that the volunteer will examine. 
The ability of volunteers to search the entire dataset for similar subjects is a methodology unique to citizen-science projects that proved highly useful; rather than proceeding through the classification task until similar-looking subjects appeared and could be added to a collection, this tool allowed project users to quickly build morphologically-similar classes of glitches that could be vetted and added to the \ac{ML} training. 
Details of the clustering algorithm and similarity search can be found in \cite{Bahaadini:2018nkh,Bahaadini:2018git,Coughlin:2019ref}.

\begin{figure}
\centering
\includegraphics[width=\linewidth]{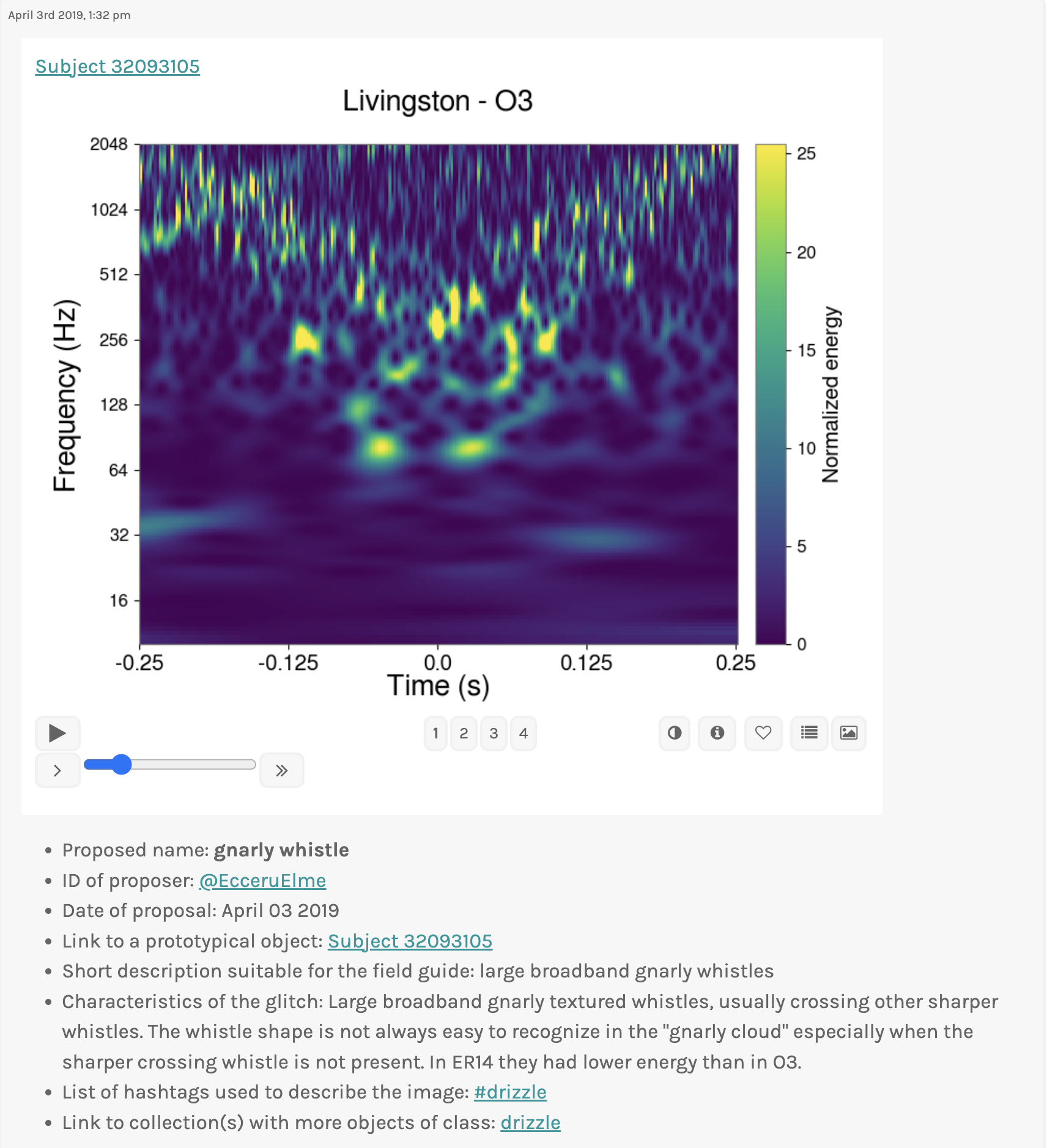}
\caption{New Glitch Proposal by Gravity Spy user EcceruElme.}
\label{fig:prop}
\end{figure}

Following the identification and curation of a new glitch class, volunteers can submit a New Glitch Proposal (Figure~\ref{fig:prop}, bottom), including a proposed name for the glitch, a description of its typical morphological features, a single exemplar or reference glitch, and their collections of similar images. 
The proposal is evaluated by a \ac{LIGO} member who assesses the robustness and usefulness of the proposed new class and then communicates whether it should be included in the list of glitch options in the classification interface.

\section{Results from Gravity Spy}\label{sec3}

Gravity Spy's impact has been three-fold: it has improved glitch mitigation in GW detectors and, thereby, the quality of GW observations; it has explored new approaches to \ac{ML} and human-computer interaction, and it has increased scientific engagement from community members. 
In the following subsections, we describe these impacts in detail. 

\subsection{Improving the Sensitivity of GW Detectors and Analyses}\label{subsec:improving_GW_detectirs}

One key investigation that Gravity Spy has enabled is monitoring the rate of glitches by glitch type. 
The disappearance of a specific glitch class can, for example, indicate that a specific mitigation strategy undertaken by the detector specialists is working~\cite{Soni:2020rbu,Davis:2021ecd}. 
Monitoring different glitch types over time provides both a high-level assessment of the detectors' state and some indication of whether a specific glitch type is responsible for increased detector noise. 
For example, Figure~5 of \cite{Glanzer:2022avx} shows the hourly glitch rates for four types of glitches at LIGO Hanford and LIGO Livingston during \ac{O3}. 

Beyond a high-level assessment of detector noise, detector-characterization experts also use Gravity Spy classifications to search for correlations between individual glitch classes and features in auxiliary channels that monitor the state of the detectors and related instruments. 
Gravity Spy classifications are easily accessible to \ac{LIGO} scientists through the internal database \texttt{LigoDV-Web}~\cite{Areeda:2016mee}. 
Investigations by detector experts, which could involve Gravity Spy classifications, are typically described in \emph{aLogs}, LIGO's records on detector functioning and data quality. 
One such example can be seen in \cite{alog:35073}, where potential correlations are identified between a sub-class of the Low-Frequency Blip class and noise in an auxiliary channel monitoring the suspension systems in \ac{LIGO} Hanford. 
Similarly, one can investigate the effects of adjusting the instruments on the rate of different classes of glitches, as exemplified in another aLog~\cite{alog:43177}.  

Gravity Spy has also had a major impact on the understanding of \ac{LIGO} noise through the identification of new glitch classes. 
For example, \ac{LIGO} scientists identified the Low-Frequency Blip class through investigations of many glitches classified by Gravity Spy as regular Blips; the subclass of Blips at lower frequencies prompted the creation of a new class in the Gravity Spy project~\cite{Soni:2021cjy}. 
Gravity Spy volunteers play a key role in this process; they can propose new glitch classes, some of which are eventually incorporated into the \ac{ML} and classifications. 
One of the first instances of this was the identification of the Paired Doves glitch class by Gravity Spy volunteers during beta testing~\cite{alog:27138,Zevin:2016qwy}. 
Paired Doves are glitches that are morphologically similar to the \ac{GW} signals expected from merging compact objects, so identifying this new class was critical for mitigating false positive \ac{GW} detections.  

Another notable example is when Gravity Spy volunteers identified the Crown glitch class, which was contemporaneously identified by \ac{LIGO} scientists as Fast Scattering~\cite{alog:44803,Soni:2021cjy}. 
These glitches were originally classified into the NOTA class or the Scattered Light class, but further investigation noted they should be their own class, separate from another sub-class of scattered light now known as {Slow Scattering}. 
The independent work of the volunteers and LIGO scientists enabled us to verify that the \ac{ML} algorithm trained using either the volunteers' Crown training set or the experts' Fast Scattering training set produced comparable results~\cite{Soni:2021cjy}, demonstrating the volunteers' capability in identifying new classes. 
This Fast Scattering/Crown class was found to be associated with ground motion and accounted for a major fraction of glitches in \ac{LIGO} Livingston during \ac{O3}~\cite{LIGOScientific:2021djp,Glanzer:2022avx}. 
The combination of quick classification by \ac{ML} and by-eye investigations by both \ac{LIGO} scientists and community members were essential to identifying and understanding this new glitch class.

Beyond direct application to characterizing glitches, Gravity Spy classifications have also been used to understand the effects of detector noise on \ac{GW} measurements and as inputs to other \ac{GW} software pipelines. 
Gravity Spy data products, including both \ac{ML} and volunteer classifications of the entire set of glitches up through \ac{O3}, are available for public use~\cite{michael_zevin_2022_5911227,coughlin_scott_2021_5649212}, with proprietary data from ongoing observing runs similarly available to LVK scientists. 
%For example, \cite{Davis:2020nyf} examines how different glitch classes from Gravity Spy may contaminate \ac{GW} searches, and many other works have used Gravity Spy classifications to curate training sets for other analyses or test the performance of alternative \ac{ML} algorithms~\citep[e.g.,][]{George:2018awu,Cabero:2020eik,Jadhav:2020oyt, Merritt:2021xwh,Yan:2022spw,Sakai:2022keo,Alvarez-Lopez:2023dmv}. 
Gravity Spy classifications have been used to examine how different glitch classes affect data analysis~\cite{Ashton:2021tvz,Macas:2022afm,Hourihane:2022doe,Heinzel:2023vkq}, for example, how they may contaminate \ac{GW} searches~\cite{Davis:2020nyf}; the development of algorithms to distinguish \ac{GW} signals from glitches~\cite{Benko:2020syv,Marianer:2020slp,Cabero:2020eik,Jadhav:2020oyt,Singh:2020yau,Abbott:2021cuf,Chaturvedi:2022suc,Choudhary:2022yje,Choudhary:2022nvs,Boudart:2022apz,Bini:2023gil,Fernandes:2023fbt,Jadhav:2023mqx,Shah:2023twc,Trovato:2023bby,Jarov:2023qpt,Alvarez-Lopez:2023dmv,Jarov2023}; the development of techniques to remove glitches from the data~\cite{Torres-Forne:2020eax,Merritt:2021xwh,Ashton:2022ztk}; investigations of potential environmental or instrumental origins of noise~\cite{Davis:2021ecd,Longo:2021avq,Colgan:2022vdd,Glanzer:2023hzf}, such as what triggers the appearance of light-scattering noise~\cite{Soni:2020rbu,Soni:2021cjy}; simulating synthetic (noisy) \ac{GW} data~\cite{Lopez:2022lkd,Powell:2022pcg,Dooney:2022arh}; and training or testing alternative \ac{ML} glitch-classification algorithms~\cite{George:2018awu,Cavaglia:2018xjq,Sankarapandian:2021qun,Sakai:2021rks,Yan:2022spw,Sakai:2022keo,Fernandes:2023fbt,Tolley:2023umc}. 
Overall, Gravity Spy classifications have been instrumental in \ac{GW} detector characterization, spanning from informal investigations to a key part of \ac{GW} data-analysis software~\cite{Davis:2022dnd}.

\subsection{Development of Machine Learning}\label{subsec:development_machine_learning}

As described earlier, Gravity Spy incorporates \acp{CNN} for machine learning-based glitch classification. 
\acp{CNN} can automatically learn hierarchical patterns and features from images to effectively distinguish between different glitch classes. 
The \ac{CNN} architecture presented in Figure~6 of \cite{Zevin:2016qwy} serves as the foundation, comprising two \ac{Conv} and max-pooling layers, followed by two \ac{FC} layers. 
The final \ac{FC} layer employs a softmax activation function, generating scores for each class based on a given set of input images. 
To create an input, spectrograms with four durations generated with different time windows ($0.5~\mathrm{s}$, $1.0~\mathrm{s}$, $2.0~\mathrm{s}$, and $4.0~\mathrm{s}$, the same that are shown to the volunteers of the project) are combined to form a square image. 
This merging process ensures the convolutional kernels effectively slide over all four durations, enabling the model to learn distinct features from both long- and short-duration glitches. 
This architecture has been trained multiple times in different contexts as shown in Table~\ref{tbl:ML_models}, as we describe below. 

The initial model was trained using a dataset of $7718$ glitches from $20$ classes observed during the \ac{O1} of Advanced \ac{LIGO}~\cite{Bahaadini:2017dqg, Zevin:2016qwy}. 
This dataset was curated and labeled via a collaboration between detector-characterization experts and citizen-science volunteers from Gravity Spy. 
%The spectrograms of each duration were resized to a dimension of $47 \times 57$, resulting in a merged input size of $94 \times 114$. 
The initial model achieved an average accuracy of $97.1\%$ on the testing set. 
Later, two new glitch classes, the 1080 Line and 1400 Ripple classes, were discovered by citizen-science volunteers during the \ac{O2} of Advanced \ac{LIGO}. 
Therefore, the model was retrained using an expanded dataset of $7932$ glitches from $22$ classes, encompassing glitches from both of the first two observing runs~\cite{Bahaadini:2018git}. 

%To accommodate the larger input dimensions of $280 \times 340$, 
The model was later improved with two additional \ac{Conv} layers and max-pooling layers before the existing two \ac{FC} layers. 
These deeper layers allowed the model to capture more intricate patterns in the glitch data. 
This retrained model achieved an accuracy of $98.2\%$ on the testing set. 
During \ac{O3}, \ac{LIGO} detector-characterization experts and the Gravity Spy volunteers identified two new glitch classes: Fast Scattering/Crown and Low-frequency Blips~\cite{Soni:2021cjy}. 
In addition, the NOTA class was removed from the glitch classes defined in the \ac{ML} model. 
This particular class is useful for volunteers to flag potential new glitch types but held limited significance for the model's classification process due to the large variety of morphological features it encompassed. 
Therefore, the model was retrained again on a dataset of $9631$ glitches from $23$ classes in \ac{O3}~\cite{Soni:2021cjy, Glanzer:2022avx}. 
%The input dimension and the model architecture remained unchanged. 
This most recent model reported training and validation accuracies of $99.9\%$ and $99.8\%$, respectively.      

While each of the \ac{ML} classifiers mentioned above demonstrated strong performance, they have some common challenges and limitations. 
First is the imbalanced distribution of glitches across classes. 
For instance, certain classes like Chirp and Wandering Line suffer from a scarcity of training samples, which can introduce bias during model training. 
To address this issue, it is recommended to incorporate class-specific evaluation metrics, such as precision and recall, when assessing the model. 
Second is the presence of noisy glitch labels. 
The training labels are derived from a combination of \ac{ML} and volunteer classification, making it likely that the model relies on partially incorrect labels during the training phase. 
Furthermore, the model architecture remains relatively simple, relying on the direct combination of a few convolutional layers. 
This simplicity imposes constraints on the model's ability to perform effective feature extraction, particularly when dealing with morphologically similar glitch classes. 
Future studies could potentially explore more advanced model architectures to further enhance the performance of the glitch classifiers. 
Finally, due to its fully supervised nature, the \ac{ML} method faces challenges in recognizing new glitch classes that fall outside the predefined classes during training, which highlights the importance of involving volunteers in the process of identifying new glitch classes.

\begin{table}
\centering
\caption{The development of \ac{ML} models for glitch classification in the Gravity Spy system. }
\label{tbl:ML_models}
\begin{tabular}{ccccc }
\toprule
Papers                                                                       & \begin{tabular}[c]{@{}c@{}}Number of\\  glitch classes\end{tabular} & \begin{tabular}[c]{@{}c@{}}Dataset\\  size\end{tabular} & \begin{tabular}[c]{@{}c@{}}Input \\ size\end{tabular}                                                                 & \begin{tabular}[c]{@{}c@{}}Model \\ architecture \end{tabular}                                                                                                     \\ \midrule
\begin{tabular}[c]{@{}c@{}}Bahaadini et al.~\cite{Bahaadini:2017dqg}\\ and Zevin et al.~\cite{Zevin:2016qwy}\end{tabular} & 20                                                                 & 7718                                                    & \begin{tabular}[c]{@{}c@{}}$94 \times 114$\\ ($47 \times 57$ \\ for each duration)\end{tabular}    & \begin{tabular}[c]{@{}c@{}}(Conv + MaxPool) $\times 2$ \\ + FC $\times 2$\end{tabular} \\ 
Bahaadini et al.~\cite{Bahaadini:2018git}                                                            & 22                                                                 & 7932                                                    & \begin{tabular}[c]{@{}c@{}}$280 \times 340$\\ ($140 \times 170$ \\ for each duration)\end{tabular} & \begin{tabular}[c]{@{}c@{}}(Conv + MaxPool) $\times 4$ \\ + FC $\times 2$\end{tabular} \\ 
\begin{tabular}[c]{@{}c@{}}Soni et al.~\cite{Soni:2021cjy}\end{tabular}   & 23                                                                 & 9631                                                    & \begin{tabular}[c]{@{}c@{}}$280 \times 340$\\ ($140 \times 170$ \\ for each duration)\end{tabular} & \begin{tabular}[c]{@{}c@{}}(Conv + MaxPool) $\times 4$ \\ + FC $\times 2$\end{tabular} \\ \botrule
\end{tabular}
\end{table}

\subsection{Engagement with Project Volunteers}\label{subsec:engaging_project_volunteers}

While a significant amount of work has studied both machine- and human-based classification schemes, we know little about how to use machine-coded data to improve human learning and performance. 
%The most critical component of Gravity Spy is the volunteers who classify the image subjects. 
To this end, we devoted substantial effort to researching how to recruit volunteers, train volunteers with the support of \ac{ML} scaffolding, and assess volunteer engagement. 

{\bf Volunteer recruitment.}
Zooniverse hosts hundreds of projects, allowing participants to be recruited from the existing user base. 
We engaged in supplemental recruitment through blog posts and announcements about \ac{GW} discoveries and empirical research on attracting and retaining participants (i.e., motivation). 
A persistent challenge in citizen-science projects is that most volunteers participate infrequently, oftentimes only once. 
Research in citizen science has focused on strategies to motivate contributions, with studies on volunteer motivation in citizen science finding that user motivation differs person-to-person and changes over time~\cite{Raddick:2009ck,Rotman:2014dw}. 

To test theories of motivation, we conducted one study focused on increasing classification volume using novelty theory~\cite{jackson2019characterizing}. 
In this experiment, we showed participants a unique message when they were the first person to see a particular glitch. 
While the theory was tested in other projects, the findings revealed novelty messages had no effect on volunteer behaviors. 
We hypothesize that other aspects of the project (e.g., gamified elements through leveling not found in other projects) might be sufficiently motivating. 
We also conducted an experiment to evaluate the efficacy of recruitment messages appealing to four types of motivations known to be salient in citizen-science projects: science learning, contribution to science, joining a community of practice, and helping scientists~\cite{lee2018appealing}. 
Messages about contributing to science resulted in more classifications; however, statements about helping scientists generated more initial interest from participants. 

{\bf Volunteer training.}
Participants of the Gravity Spy project may have little classification task experience, so training volunteers is crucial for ensuring data quality. 
The Zooniverse platform supports training through text- and image-based tutorials. 
We conducted several investigations to gauge the efficacy of our training and learning regimens. 
One such study was conducting an online A/B field experiment to evaluate the training procedure described in Section~\ref{subsec:2.2}~\cite{jackson2020teaching}: Group A received all glitch classes (with a wide range \ac{ML} confidence) without training when they started the project, whereas Group B received the default scaffolded training through the workflow levels that increase in complexity and difficulty as they proceed through the project. 
As anticipated, we found that volunteers who received the training were more accurate in their classifications (as indicated by agreement on gold-standard data and with the eventual consensus decision on the glitch) and contributed significantly more classifications to the project. 

Another study used digital trace data (produced as a by-product of interactions with computer systems) to evaluate how volunteers use learning resources when given feedback about their classification (e.g., ``You answered Blip, but our experts classified this image as Whistle''). 
The results demonstrated that authoritative knowledge provided by the project scientists improved learning during early workflows. 
However, as the challenge increases in advanced workflows, volunteers rely on resources constructed by the community (e.g., discussion boards) to learn to identify glitches more accurately~\cite{jackson2020shifting}. 
The results of our investigation into the learning process have informed design decisions for integrating scaffolded tasks in subsequent citizen science initiatives.
Our findings outline optimal strategies to blend human and \ac{ML} schemes to train volunteers and identify the temporal importance of learning resources. 
Additionally, this research underscores the significance of learning resources generated by volunteers in the absence of expert guidance. 

{\bf Volunteer engagement.}
Much research has been conducted about people engaging in virtual citizen science projects. It tends to demonstrate that many volunteers contribute once or infrequently while a handful of volunteers perform the majority of work and limit their engagement to image classification \cite{Sauermann:2015kp, Rohden:2019hw}. 
This observation is corroborated by a Gini coefficient of 0.85, indicative of a high level of inequality in participation as found in other citizen science projects \cite{spiers2019everyone}. 
In Gravity Spy, on average, a volunteer makes 235 classifications (median = 2) before dropping out, and less than 12\% of volunteers have contributed to the project's discussion boards.  
While engagement is skewed, a small cadre of highly motivated and engaged volunteers is active on the site. 
Our analysis of digital trace data shows how volunteers engage beyond submitting classifications, engaging with discussion boards and other project infrastructure. 
Our research describes how volunteers share new knowledge by linking external resources (e.g., arXiv preprints) and sharing results of independent investigations speculating about the causes of glitches in the data \cite{jackson2018folksonomies, ekstrom2021tracing}. 
As noted above, many glitch class proposals are the product of intense data curation and evaluation on the discussion boards. 
We find both independent and collaborative interactions~\cite{harandi2021occasional}. 
Overall, our research on engagement demonstrates the ability of volunteers to engage in more complex work when provided the means to do so. 

{\bf Volunteer contributions.}
Volunteer efforts have been crucial to Gravity Spy. 
At present, over $7.4$ million classifications have been performed by project volunteers, with more than $32,000$ registered users (and many more users that did not register with a Zooniverse account) contributing to these classifications. 
Volunteers have drafted over $30$ new glitch proposals: in \ac{O3}, new glitch classes from these proposals were included in the glitch classification interface and used in retraining the \ac{ML} model~\cite{Soni:2021cjy,Glanzer:2022avx}.%
%\footnote{Examples of new O3 glitch classes \href{https://blog.gravityspy.org/2019/05/07/new-o3-glitch-options/}{blog.gravityspy.org/2019/05/07/new-o3-glitch-options/}}

\section{Challenges and Future Considerations}\label{subsec:challenges}

Gravity Spy has faced several challenges that can be traced back to the intersection between the science team’s and volunteers’ work: (1) science team--volunteer engagement; (2) mismatching temporal rhythms and divided priorities across the science team and volunteers; (3) retraining of \ac{ML} models, and (4) the constant evolution of the \ac{GW} detectors. 

First, the science team often lacked the capacity for sustained volunteer interactions, a challenge for many citizen science projects \cite{doi:10.1080/21548455.2020.1719288}.
In Gravity Spy, this was evidenced through bottle necks in approving new glitch class proposals and questions posted to discussion boards. 
Glitch proposals required extensive efforts by the science team to evaluate the veracity of the proposed glitch in the data stream, whether proposed glitch classes are still impacting \ac{GW} detectors and relevant to detector characterization science, and, when proposals were rejected, justifying the decision to the volunteers. 
For some proposals, many months can pass between an initial glitch proposal by a volunteer and a resolution to the proposal by the science team.\footnote{Gravity Spy Talk: Gnarly Whistle Glitch Proposal \href{https://www.zooniverse.org/projects/zooniverse/gravity-spy/talk/762/951832}{www.zooniverse.org/projects/zooniverse/gravity-spy/talk/762/951832}, and Pizzicato Glitch Proposal \href{https://www.zooniverse.org/projects/zooniverse/gravity-spy/talk/762/935664}{www.zooniverse.org/projects/zooniverse/gravity-spy/talk/762/935664}.} 
In evaluating glitch proposals, we trained physics undergraduate and graduate students to review new glitch class proposals. 
However, student schedules quickly fill up, or they move on to other projects, so solely relying on students for glitch proposal review is not a long-term solution. 
Meanwhile, some moderators know as much or more about the detectors as students and even members of the science team, but lack formal training, easy access to experienced science team members, and password-protected \ac{LIGO} resources. 
In such instances, we have asked moderators to tag science team members who they believe have expertise in the question. 
Another challenge is that volunteers do not always realize that seemingly simple questions posted to discussion boards may take hours to address satisfactorily.  

We have experimented with different methods to triage questions to the relevant experts and enrolled detector-characterization group members (not already a part of the Gravity Spy research team) to be on call to respond to questions on the discussion boards. 
We quickly discovered that many questions pertained to the project infrastructure (e.g., debugging the promotion algorithm) or similar issues that could only be handled and addressed by a small number of people or a single person on the project team.

In the latest version of the project (see Section~\ref{subsec:gs2}), we strive to make volunteers more self reliant by giving them access to more background knowledge about the detectors when needed (the same knowledge science teams use to evaluate glitch class proposals). 
One element of this solution involves building a wiki that gradually expands volunteers' access to expert knowledge about the detectors \cite{crowston2023design, corieri2023advanced}.

Second, the temporal rhythms and priorities guiding science team members' and volunteers' work are not always consistent. 
For instance, the investigations by the detector-characterization group and volunteers do not always align. 
While the detector-characterization team works on priorities for an entire observing run, which may involve time-sensitive investigations and fixes to the instruments, these may occur before volunteers can identify a new glitch, given the time required to build collections and develop solid proposals about a new glitch class. 
In contrast, volunteers may submit new glitch class proposals involving glitches unknown to the detector-characterization group (e.g., the Pizzicato glitch). 
To solve this issue, we are considering a real-time citizen science format where volunteers would work on urgent problems and new data that call for quick turnarounds. 

Similarly, the science team often needs nearly real-time classification of glitches. 
As a result, they have come to rely on the automated \ac{ML} classifications, rather than waiting for the theoretically better human-classified results. 
Our realization of this dynamic has led to a refocusing of the project: more quickly retiring images rather than expending volunteer effort to refine them, using the volunteer results for model retraining rather than as immediate feedback to \ac{LIGO} scientists, and emphasizing the role of volunteers in the discovery and characterization of novel glitch classes. 

The third major challenge in the Gravity Spy project has been that retraining the \ac{ML} model based on volunteer classifications has taken longer than initially planned. 
The issue has multiple elements. 
The dataset containing volunteer classifications has impurities, and we have had to experiment with different ways to assess its accuracy. 
Rare glitch classes have few examples, making the integration of them into a retrained model difficult. 
From a \ac{ML} perspective, dealing with four versions of the same glitch images in different time durations can also be challenging. 
While it is helpful for the volunteers to move between different time durations, the \ac{ML} can sometimes regard some time durations as more important than others, leading to potential classification problems.  

Fourth, \ac{GW} detectors continually evolve~\cite{LIGOScientific:2016emj,Abbott:2016xvh,LIGOScientific:2017bnn,Buikema:2020dlj}; with it, new glitch classes emerge, and known classes disappear or become less prevalent. 
One cannot leave the \ac{ML} model unattended for long. 
At the beginning of \ac{O4}, the volunteers and, soon after, the science team found that the \ac{ML} model was confidently mislabeling instances of a novel glitch as a Whistle glitch, which is one of the two glitch classes presented to volunteers in their first workflow. 
During this initial portion of \ac{O4}, the detectors did not produce many Whistles, even though they used to be common enough that Gravity Spy used them for training purposes in the initial workflow. 
As such, we are considering a replacement for the Whistle at the introductory volunteer level as we retrain the \ac{ML} model to distinguish Whistles from the new glitch. 

The constant evolution of the detectors emphasizes the importance of clearly distinguishing the architecture of the \ac{ML} model from the different retrained versions of a specific model. 
We have had to carefully track how the model changes as one experiments with new, more effective architectures. 
Equally important, we have had to verify the provenance of the different versions of a model as it gets retrained on new classification data, some of which might include new glitches. 
This is not always easy in a distributed science team with many stakeholders interested in the continuous improvement of the model. 

Finally, making all the improvements deemed necessary is not always possible. 
For instance, we have yet to complete a volunteer performance assessment system that moves beyond relying on gold-standard data. 
The current \ac{ML} model would guide the presentation of image classification tasks to newcomers to help them learn how to do the task more quickly while contributing to the project's work. 
We designed but never implemented a Bayesian model that would both estimate volunteers' ability and decide the classification of an image. 
The reason for not implementing this scheme was due to occasional high-confidence misclassifications from the \ac{ML} model, which, when the volunteers made an alternative selection, would hinder their ability to progress through the project workflows. 
A simulation of the model applied to volunteer promotion and image retirement suggests that the model would require fewer classifications than the current system and, in the process, save volunteers’ time \cite{crowston2019knowledge}. 

The challenges that Gravity Spy faces need to be understood in the context of its successes. 
The Gravity Spy project has led to substantial innovations in combining machine learning  and volunteer based classification schemes, and demonstrated how the facilitation of a symbiotic relationship between these methodologies leads to significant improvement in classification and characterization tasks. 
The Gravity Spy classifications and associated models have benefited \ac{GW} detector characterization, and the project has engaged more than $30$ thousand volunteers in the scientific process. 
This success was tied to the substantial user base and promotion that Zooniverse has developed over the past $15$ years, as well as the public interest in the growing field of \ac{GW} science since the start of \ac{O1} in 2015. 
The project also benefits from an active group of advanced volunteers. 
Led by a dedicated team of volunteer moderators, Gravity Spy continues to discover, collect, characterize, and name new glitch classes. 
The volunteer moderators have also played a critical role in responding to the many questions of other users in the project, a feat that would not have been possible by the science team alone. 
The scaffolded training where volunteers gradually level up has greatly improved volunteer engagement and retention~\cite{jackson2020teaching}, and is being implemented across other Zooniverse projects. 
To a certain extent, the success of the Gravity Spy classification model and the high volunteer engagement has removed the urgency of addressing some the outstanding issues in the project itself. 

\section{Gravity Spy 2.0}\label{subsec:gs2}

The success and challenges of Gravity Spy have spurred us to explore how volunteers can engage in more complicated analysis of \ac{LIGO} data that would help detector-characterization scientists identify the causes of glitches and isolate these signals in the data stream. 
Along with the main strain channel that is sensitive to \acp{GW}, the \ac{LIGO} detectors record more than $200,000$ auxiliary channels of data per detector from a diverse set of sensors that continuously measure every aspect of the detectors and their environment (e.g., equipment functioning, activation of components, seismic activity, or weather)~\cite{TheLIGOScientific:2016zmo,Nguyen:2021ybi}. 
To explore the cause of glitches (i.e., what is happening in the detector or the environment that causes particular glitches), detector-characterization scientists conduct investigations using the temporal correlation of these auxiliary channels and the main \ac{GW} channel. 
Using these data, researchers further isolate noise signals in the data stream, often analyzing output data using statistical software and conducting visual inspections of \ac{GW} and auxiliary channel glitch images. 

In 2020, the Gravity Spy collaboration began work funded by a new multi-year National Science Foundation grant (HCC 2106865) focused on developing the capacity for amateur volunteers to conduct similar activities and develop causal insights about glitches.  
This project, known as \textit{Gravity Spy 2.0}, is currently live as part of the broader Gravity Spy infrastructure. 
It is an extension of our efforts in the initial Gravity Spy and is designed to help volunteers develop their knowledge of the data over time by progressing through three stages (also Phases of the grant). 
\vspace{3pt}
\begin{enumerate}[label={\bf Stage \arabic*:},leftmargin=3.5\parindent]
    \item Volunteers will compare glitches in the main \ac{GW} channel to glitches in auxiliary channels. 
    At first, we provide users with $3$ auxiliary channels per glitch subject, totaling in $\sim~20$ distinct auxiliary channels being used across the glitch classes considered in the first phase of the project. 
    Human input is valuable because there could be morphological similarities between the glitches in the channels that point to some association. 
    However, there is no simple rule to determine which channels are relevant or how the channels are coupled. 
    Engagement with the data and the channel ontology in this phase will also support volunteers in learning about the auxiliary channels and their potential relationship to a glitch in the \ac{GW} channel. 
    A similar approach has been tried using data from the Virgo detector in the GWitchHunters project~\cite{Razzano:2022lgg}.\footnote{GWitchHunters Project \href{https://www.zooniverse.org/projects/reinforce/gwitchhunters}{www.zooniverse.org/projects/reinforce/gwitchhunters}.}
    \item The primary task will shift from individual glitches to collections, building on the filtered dataset of glitches and auxiliary channels created in Stage~1. 
    Volunteers are expected to (1) create collections of glitches that might represent a novel glitch class with hypothesized common causality, not just similar appearance; (2) search for relations between those glitches and groups of auxiliary channels over time, and (3) note possible causes behind the correlated glitches and auxiliary channels.
    \item Advanced volunteers will analyze the network compiled at Stage~2 to identify the root causes of glitches. 
    A supporting activity is adding paths among auxiliary channels, e.g., by examining patterns that appear across multiple classes of glitches. 
    For instance, if several channels are involved simultaneously in many classes of glitches, they are likely interrelated.
\end{enumerate}
\vspace{3pt}
In each stage, we will use novel interfaces and tools to support volunteer tasks. 
Our research adopts a \emph{human-centered design} approach involving volunteers and detector-characterization scientists in developing Gravity Spy 2.0. 

Development of Gravity Spy 2.0 and research into how best to use its output is ongoing. 
Discussions with detector-characterization scientists are needed to understand and translate their work making causal inferences. 
To that end, we need to understand what background knowledge about the glitches and the detector is necessary and how to structure the tasks. 
On the volunteer side, we need to know how best to align the system support with volunteers' current knowledge and capabilities. 
To that end, we recently completed interviews and design critiques with long-time Gravity Spy volunteers. 
During the user studies, we gained insights into volunteers' glitch classification practices, which will inform how we design the three stages. 

Much like the original Gravity Spy, the output of this investigation will contribute new knowledge about cutting-edge citizen-science techniques that involve amateur volunteers in more advanced scientific tasks. 

\section{Citizen Science in Physics and Beyond}\label{sec:conclusions}

Gravity Spy's use and advancement of strategies that combine human and machine effort have produced an important blueprint for future citizen-science projects in physics and a broad range of other research domains. 
The capabilities of human crowdsourcing alone cannot keep up with the demands of ever-growing data volumes, even when assuming continued access to large communities of volunteers. 
However, the value of human input is unquestionably high, particularly in applications such as out-of-sample classification, serendipitous discovery, and other advanced analyses. 
Therefore, the continued development and application of hybrid machine--human workflows is essential for unlocking the full potential of many datasets.

Gravity Spy has demonstrated paths forward for capitalizing on the capabilities and strengths of human crowdsourcing. 
First, the \ac{ML}-informed assignment of human effort in Gravity Spy, where subjects with high machine confidence are tasked to beginners and retired quickly and subjects with uncertain or unidentified machine classifications are assigned to experienced volunteers, is a model applicable to many projects. 
If \ac{ML} models can expedite the classification of easy and routine data, that leaves greater effort for humans to contribute in more meaningful and advanced ways. 
Moreover, if projects successfully combine machine and human classifications for use in improving and refining the machine model, the resulting classification quality and system efficiency can be improved and optimized. 
The training set augmentation and model retraining strategy of Gravity Spy or the active learning strategy employed by a recent iteration of the Galaxy Zoo project~\cite{Walmsley2020}, show the benefits of this human--machine interplay and its applicability in multiple research domains.

Second, Gravity Spy affirms a broader trend across Zooniverse citizen-science projects that a subset of volunteers is willing to contribute in advanced ways that go well beyond the project's primary classification task. 
The investigation and proposal of new glitch classes and the use of the project's Similarity Search tool show the willingness of certain volunteers to perform information synthesis activities, which augment the typical, more narrow scope of machine classifiers. 
This behavior mirrors that seen in other projects, such as volunteers performing follow-up analysis of previously unknown galaxy sub-types via the Galaxy Zoo project (e.g., Green Pea galaxies~\cite{Cardamone:2009me}) or the use of publicly available external data and software to characterize exoplanets as part of the Planet Hunters TESS project~\cite{Eisner2021}. 
A common thread among these examples is the ability for volunteers to access data and tools that allow additional contributions towards the project science goals, and the resulting reward in the form of new findings to project teams that facilitate and support these volunteer activities. 

Third, the patterns of interaction associated with Gravity Spy's glitch proposals demonstrate the importance of setting up effective processes for interaction and feedback between research teams and volunteers, especially those carrying out advanced work. 
While ambitious volunteers may take it upon themselves to reach out and communicate with the project team using basic means (email, Zooniverse Talk boards, etc.), both the research team and volunteers often benefit from creating structures and processes to facilitate useful and beneficial interactions. 
This conclusion motivates a recommendation that citizen-science projects craft a plan for volunteer interaction that includes tools for both communication and analysis, especially to encourage more independent forms of volunteer contribution. 
We highlight that the Gravity Spy team's interdisciplinary composition was an important asset in assembling a successful plan for volunteer engagement and advocate that research teams that include members spanning a broad range of project-enabling skills will better position their projects for potential success.

% Concluding Paragraph?
%\lcj{This is placeholder text - please improve!}
In summary, Gravity Spy has pioneered how \ac{ML} approaches and citizen science can be synthesized across various domains to produce useful results. 
\ac{GW} science has been supported in the process through large-scale classification and characterization of glitch classes and identification of new glitch morphologies, thereby enabling the identification of glitch origins and the removal of glitches from the \ac{GW} datastream or eliminating the root cause of glitches from the detectors entirely. 
The next phase of the Gravity Spy project and the continued flow of new observations from \ac{LIGO}, Virgo and KAGRA show that this project will continue to meaningfully contribute to scientific discovery for years to come. 

% QUESTIONS FROM THE EDITOR: What do you estimate as the cost/benefit ratio to LIGO? To the citizen participants? Have you had any ideas about theoretical as opposed to observational campaigns? A list (as technical and concrete as possible) of recommendations for related citizen-science projects that could/should be used in Relativity/Astrophysics but also encompass crowdsourcing for scientific projects in general. From the beginning, the point is to define how to set up scientifically relevant citizen-scientist sourcing, so some of the more “administrative” details of creating the system might be valuable.

% Original Pitch - kept here, but queue for removal?
%How we’re using what we learned from Gravity Spy to build/improve Gravity Spy 2.0
%Leverage the diverse skillsets of volunteers. Some are good at pattern recognition and may only want to engage in this work; others are interested in more advanced work uncovering causal relationships between different noise signals and their appearance in the detectors. 
%\cplb{Add emphasis on volunteers being able to make discoveries if provided with the appropriate framework?}
%Interdisciplinary teams - a group of astrophysicists, computer scientists, and social scientists are working to make this happen. 
%Algorithm provenance - NW issue with naming and handing off \ac{ML} to other groups. It’s also an interesting discussion about \ac{ML} model building and its different parts. But that might be secondary to the discussion about science use.

\backmatter

\bmhead{Acknowledgments}
The authors first and foremost are indebted to the citizen-science volunteers of the Gravity Spy project, whose dedicated work allowed for this ambitious project to become a reality. 
The authors also thank Laura Nuttall, Lynn Cominsky, and the anonymous referees for comments that improved this manuscript. 
This publication uses data generated via the Zooniverse.org platform, development of which is funded by generous support, including a Global Impact Award from Google, and by a grant from the Alfred P.\ Sloan Foundation.
Gravity Spy is partly supported by the National Science Foundation (NSF) awards INSPIRE 1547880,  IIS-2107334, 2106882, 2106896, 2106865, 2107334 and  PHY-1912648. 
Support for MZ was provided by NASA through the NASA Hubble Fellowship grant HST-HF2-51474.001-A awarded by the Space Telescope Science Institute, which is operated by the Association of Universities for Research in Astronomy, Incorporated, under NASA contract NAS5-26555. 
ZD is supported by the CIERA Board of Visitors Research Professorship.  
ZD, VK, SB, and JS acknowledge support from PHY-2207945.
YW, VK, SB, and RH were also supported by IIS-2107334. 
VK is grateful for support from a Guggenheim Fellowship, from CIFAR as a Senior Fellow, and
from Northwestern University, including the Daniel I.
Linzer Distinguished University Professorship fund.
CPLB acknowledges past support from the CIERA Board of Visitors Research Professorship, and current support from Science and Technology Facilities Council (STFC) grant ST/V005634/1. 
DD is supported by the NSF as a part of the LIGO Laboratory. 
SS is supported by NSF Award PHY-1764464 to LIGO Laboratory. 
This material is based upon work supported by NSF's LIGO Laboratory, a major facility fully funded by the National Science Foundation. 
This work used computing resources at CIERA funded by NSF grant PHY-1726951, and the computational resources and staff contributions provided for the Quest high performance computing facility at Northwestern University, which is jointly supported by the Office of the Provost, the Office for Research, and Northwestern University Information Technology.
The authors are grateful for computational resources provided by the LIGO Laboratory and supported by National Science Foundation Grants PHY-0757058 and PHY-0823459. 
This document has been assigned LIGO document number \href{https://dcc.ligo.org/LIGO-P2300256}{LIGO-P2300256}. 
%Glitch classification data from the Gravity Spy project are openly available from Zenodo~\cite{coughlin_scott_2021_5649212,michael_zevin_2022_5911227}.

\bmhead{Data Availability Statement}
The Gravity Spy datasets generated and analyzed during the current study, including machine learning and volunteer classification information, are available in Zenodo repositories~\cite{coughlin_scott_2021_5649212,michael_zevin_2022_5911227}. 
Datasets generated for Gravity Spy 2.0 are available from the corresponding author on reasonable request. 

\bibliography{sn-bibliography}

\end{document}